\documentclass[
  superscriptaddress,
  amsmath,amssymb,
  aps,
  reprint,
  prb,
  showkeys,
  longbibliography
]{revtex4-1}

\usepackage{graphicx}  
\usepackage{siunitx}
\usepackage{caption}
\usepackage{subcaption}
\usepackage{dcolumn}
\usepackage{bm}
\usepackage{physics}
\usepackage[normalem]{ulem}
\usepackage[version=4]{mhchem}
\usepackage[colorlinks=true, citecolor=blue, linkcolor=blue, urlcolor=blue]{hyperref}
\usepackage[capitalize]{cleveref}
\usepackage{comment}
\usepackage{xspace}

\newcommand{\tc}{$T_{\text{c}}$\xspace}
\newcommand{\rs}{$R_{\text{s}}$\xspace}
\newcommand{\isat}{$I_{\text{sat}}$\xspace}
\newcommand{\isatnorm}{$I_{\text{sat,norm}}$\xspace}
\newcommand{\idet}{$I_{\text{det}}$\xspace}
\newcommand{\idetnorm}{$I_{\text{det,norm}}$\xspace}
\newcommand{\isw}{$I_{\text{sw}}$\xspace}
\newcommand{\iret}{$I_{\text{ret}}$\xspace}
\newcommand{\csub}{interfacial thermal boundary conductance\xspace}
\newcommand{\mosi}[2]{Mo$_{\text{#1}}$Si$_{\text{#2}}$\xspace}

\newcommand{\figreflet}[2]{\hyperref[#1]{Figure~\ref*{#1}#2}}
\DeclareUnicodeCharacter{2212}{\ensuremath{-}} 
\DeclareUnicodeCharacter{03BC}{\ensuremath{\mu}} 
\DeclareUnicodeCharacter{00B5}{\ensuremath{\mu}} 
\begin{document}
\title{Enhancing the Sensitivity of Amorphous MoSi SNSPDs by Helium Ion Irradiation}
\date{\today}

\author{Fabian Wietschorke}
\thanks{These two authors contributed equally}
\email{fabian.wietschorke@tum.de}
\affiliation{Walter Schottky Institut, Technische Universit{\"a}t M{\"u}nchen, 85748 Garching, Germany}
\affiliation{TUM School of Computation, Information and Technology, Technische Universit{\"a}t M{\"u}nchen, 80333 M{\"u}nchen, Germany}

\author{Stefanie Grotowski}
\thanks{These two authors contributed equally}
\email{fabian.wietschorke@tum.de}
\affiliation{Walter Schottky Institut, Technische Universit{\"a}t M{\"u}nchen, 85748 Garching, Germany}
\affiliation{TUM School of Natural Sciences, Technische Universit{\"a}t M{\"u}nchen, 85748 Garching, Germany}

\author{Jonas Wiebe}
\affiliation{Walter Schottky Institut, Technische Universit{\"a}t M{\"u}nchen, 85748 Garching, Germany}
\affiliation{TUM School of Computation, Information and Technology, Technische Universit{\"a}t M{\"u}nchen, 80333 M{\"u}nchen, Germany}

\author{Julian Sickinger}
\affiliation{Walter Schottky Institut, Technische Universit{\"a}t M{\"u}nchen, 85748 Garching, Germany}
\affiliation{TUM School of Computation, Information and Technology, Technische Universit{\"a}t M{\"u}nchen, 80333 M{\"u}nchen, Germany}

\author{Matthias Althammer}
\affiliation{Walther-Mei{\ss}ner-Institut, Bayerische Akademie der Wissenschaften, 85748 Garching, Germany}
\affiliation{Munich Center of Quantum Science and Technology (MCQST), 80799 M{\"u}nchen, Germany}
\author{Rudolf Gross}
\affiliation{Walther-Mei{\ss}ner-Institut, Bayerische Akademie der Wissenschaften, 85748 Garching, Germany}
\affiliation{Munich Center of Quantum Science and Technology (MCQST), 80799 M{\"u}nchen, Germany}

\author{Jonathan J. Finley}
\affiliation{Walter Schottky Institut, Technische Universit{\"a}t M{\"u}nchen, 85748 Garching, Germany}
\affiliation{TUM School of Natural Sciences, Technische Universit{\"a}t M{\"u}nchen, 85748 Garching, Germany}
\affiliation{Munich Center of Quantum Science and Technology (MCQST), 80799 M{\"u}nchen, Germany}

\author{Kai M{\"u}ller}
\affiliation{Walter Schottky Institut, Technische Universit{\"a}t M{\"u}nchen, 85748 Garching, Germany}
\affiliation{TUM School of Computation, Information and Technology, Technische Universit{\"a}t M{\"u}nchen, 80333 M{\"u}nchen, Germany}
\affiliation{Munich Center of Quantum Science and Technology (MCQST), 80799 M{\"u}nchen, Germany}

\begin{abstract}
    Helium-ion irradiation has emerged as an effective post-processing technique for superconducting nanowire single-photon detectors (SNSPDs), but most studies have focused on polycrystalline superconductors, while the effects on amorphous materials remain less understood. This study investigates He-ion irradiation of amorphous \ce{MoSi} SNSPDs with varying He fluences and examines the impact on detector sensitivity, transport properties and material composition of the thin films.
    The results reveal that intrinsic photon sensitivity increases with increasing He-ion fluence, while the switching current decreases substantially. Analysis of the irradiated thin films shows a significant reduction in the transition temperature and a mild increase in sheet resistance with increasing ion fluence, attributed to a decrease in the electrically active film thickness and stoichiometry changes in the superconductor. Stopping and Range of Ions in Matter (SRIM) simulations indicate irradiation-induced intermixing at the interfaces between the \ce{MoSi} layer and both the capping and substrate layers. Based on these findings, a phenomenological model is proposed that relates the observed intermixing to changes in the effective superconducting thickness. These results provide insights into the mechanisms behind irradiation-induced changes in the superconducting material. He-ion irradiation can thus enhance the sensitivity of amorphous silicide-based SNSPDs, with potential for extending detection towards longer wavelengths.
\end{abstract}


\maketitle

\section{Introduction}
Superconducting Nanowire Single-Photon Detectors (SNSPDs) \cite{Goltsman2001} are a leading technology for quantum technologies \cite{Nerenberg2025, Eaton2023, Cacciapuoti2020} as they unite high system detection efficiency \cite{Chang2021}, ultra-low dark count rate \cite{Shibata2013} and high timing resolution \cite{Korzh2020}. 
SNSPDs offer a high detection efficiency across a wide range of wavelengths \cite{Haas2007}, ranging from the UV \cite{Wollman2017}, through the visible \cite{Yamashita2014} to the NIR-regime \cite{Luskin2023}. 
SNSPDs based on amorphous materials, such as MoSi \cite{Korneeva2014, Verma2015} or WSi \cite{Verma2014, Zhang2016} are especially suited for single-photon detection in the NIR regime, due to their smaller superconducting energy gap compared to Nb-based superconductors. Also, the material composition \cite{Zichi2019, Grotowski2026, Zhang2016} is a powerful tuning parameter to improve the detector sensitivity for long wavelengths. Helium ion irradiation has emerged in recent years as a versatile post-processing technique for tuning SNSPD performance. Alongside the demonstrated enhancement of detector sensitivity, growing experimental and theoretical work has improved the understanding of the underlying changes in superconducting, structural, and thermal properties \cite{Strohauer2023, Zhang2019, Wang2026, Hong2025}. This increasing understanding has also enabled more targeted applications.
The enhancement originates from a combination of multiple irradiation-induced effects \cite{Strohauer2025}. 
Ion irradiation can enhance the detection sensitivity of SNSPDs through two main mechanisms. First, irradiation introduces defects into the superconducting film and modifies its transport and superconducting properties, including \tc and $R_\mathrm{s}$. The resulting reduction of the superconducting energy gap lowers the energy required for photon detection and can extend the sensitivity towards longer wavelengths \cite{Strohauer2023, Zhang2019, Wang2026}.
Second, irradiation can modify the film-substrate interface and the substrate itself. Defect formation, amorphization, and helium-bubble formation can reduce the thermal coupling to the substrate and slow down hotspot cooling. A larger fraction of the deposited photon energy therefore remains in the superconducting film for longer, which can further enhance the detection sensitivity \cite{Zhang2019, Xu2023, Hong2025}. Additional structural effects induced by irradiation may also contribute, although their role is less well established.

So far, mainly the irradiation of polycrystalline materials such as NbTiN \cite{Strohauer2023} and NbN \cite{Zhang2019} has been studied. Irradiation of silicide-based detectors is therefore interesting for two reasons. On one hand, it may provide a way to extend the already high sensitivity of silicide-based detectors, which have demonstrated photon detection up to \SI{29}{\micro\meter} \cite{Hampel2026, Taylor2023}. On the other hand, it is interesting to investigate whether a defect-generation technique such as He-ion irradiation can significantly modify an amorphous superconductor that lacks long-range crystalline order. This can provide valuable insight into the mechanisms responsible for the irradiation-induced enhancement of detector performance.

In this work, we first show how the sensitivity of detectors made from amorphous materials is enhanced through He-ion irradiation and discuss the influence of He-ion irradiation on MoSi thin films. 
We use two thin films with different stoichiometries and analyze the transition temperature \tc and sheet resistance \rs to discuss the changes of the switching current and sensitivity of the SNSPDs. 
Lastly, we use SRIM simulations to investigate the recoil and intermixing effects in the thin film stack upon ion irradiation and how this can alter the thin film properties. 

\section{Experimental}
\subsection{Fabrication}
The thin films are grown by magnetron co-sputtering inside an ultra-high vacuum chamber, with the base pressure kept below \SI{1.4e-8}{\milli \bar} prior to deposition. 
The molybdenum (Mo) and silicon (Si) targets, each 2-inch in diameter, are driven independently: the \ce{Si} target runs at a fixed RF power of \SI{75}{\watt}, while the \ce{Mo} target is supplied with a DC power between \SI{35}{\watt} and \SI{66}{\watt}. 
Deposition occurs in an argon atmosphere at a working pressure of \SI{3.1e-3}{\milli \bar}. 
Two thin films, Film~A and Film~B, of \SI{4.9}{\nano \meter} and \SI{5.1}{\nano \meter} thickness are deposited onto \ce{Si} substrates with a \SI{130}{\nano \meter} thermally oxidized SiO$_2$ surface layer. 
An amorphous \ce{Si} capping layer of \SI{4}{\nano \meter} thickness protects the film from oxidation. 
During deposition the layer thickness is tracked via a quartz crystal which determines the deposition rate.
Subsequently, we determine the resulting thickness by x-ray reflectivity (XRR). 

The SNSPD fabrication begins with defining the nanowire pattern in \textit{AR-N 7520.073} resist by electron beam lithography, which is then transferred into the superconducting film through \ce{CF4}-based reactive ion etching. Electrical contacts consisting of evaporated Ti/Au pads are added via optical lithography. Each device has a superconducting nanowire width of \SI{100}{\nano \meter}. 
Furthermore, to determine the transport properties, we pattern cloverleaf structures used for four point measurements \cite{vanderPauw1958, Miccoli2015, Strohauer2023} by optical lithography. Etching and contact pad evaporation are identical to the detectors. 
The post-process irradiation is performed in a He-ion microscope (\textit{Zeiss Orion Nanofab}) with an acceleration voltage of \SI{30}{\kilo \volt}. 
The resulting He-ion fluence is varied between \SI{100}{ions \per \nano \meter^2} and \SI{2000}{ions \per \nano \meter^2}. 
For the cloverleaf the full structure is irradiated, whereas for the SNSPD only a partial irradiation is performed. By irradiating only the straight segments of the SNSPD without the 180 degree turns this technique counteracts current crowding induced effects, as described in detail by Strohauer\textit{et al.} \cite{Strohauer2025a}. 

\subsection{Measurement}
Samples are characterized at a base temperature of \SI{1}{\kelvin} in an  ADR cryostat (\textit{Kiutra}). 
For the electrical measurements, such as the switching current, a DC current is applied, provided by a series resistor (\SI{50}{\kilo \ohm}) and a voltage source. 
For count rate measurements, an additional \SI{50}{\ohm} shunt resistor is placed in parallel, to prevent latching and stabilize the bias current. 
The RF output signal is separated from the DC circuit by a bias tee and amplified in two stages: 
first, at \SI{3}{\kelvin} (\textit{Cosmic Microwave Technology, CITLF3}) and second, at room temperature (\textit{RF Bay, LNA2000}). 
Lastly, the output signal is recorded using a counting device (\textit{Swabian Instruments, TimeTaggerX}). The devices are illuminated via a free space CW laser at \SI{780}{\nm}.

The transport properties are obtained via the van-der-Pauw method, a specialized four point measurement scheme that uses a symmetrical structure, resembling a cloverleaf (CL) \cite{vanderPauw1958, Miccoli2015, Strohauer2023}. 
The sheet resistance \rs is evaluated at room temperature and \SI{20}{\kelvin}, \tc is extracted from the resistance measurements recorded from \SI{20}{\kelvin} to below \SI{2}{\kelvin} and is defined as the temperature at which the resistance reaches 50\% of its normal-state value at \SI{20}{\kelvin}.
The quasiparticle diffusivity $D$ is derived from measurements of the superconducting transition under applied magnetic fields ranging from \SI{-0.1}{\tesla} to \SI{1.1}{\tesla}, following the approach of Strohauer \textit{et al.} \cite{Strohauer2023}.

\section{Results}
\begin{figure}[t]
    \centering
    \includegraphics[width=\columnwidth]{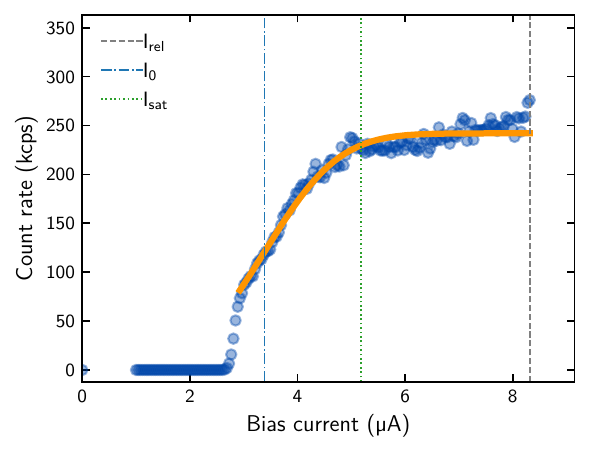}
    \caption{Count rate versus bias current for the detector irradiated with \SI{1000}{ions\per\nano\meter^2}. The detector exhibits a pronounced saturation plateau above $0.95\cdot CR_\mathrm{max}$, starting at \isat. The inflection point $I_\mathrm{0}$ corresponds to $0.5\cdot CR_\mathrm{max}$, while $I_\mathrm{rel}$ marks the onset of relaxation oscillations. Below \SI{2.8}{\micro\ampere}, the measured count rate is reduced by the high counter threshold. Dark counts are omitted, as the dark count rate remains below \SI{2}{cps} over the entire displayed range.}
    \label{fig:countrate_plot}
\end{figure}
Previous works have shown that He-ion irradiation of NbTiN-SNSPDs enhances sensitivity by changing the structural properties \cite{Strohauer2025, Hong2025} and inducing defects in the superconducting thin film \cite{Sarkar2026, Strohauer2023, Zhang2019}. 
The SNSPDs investigated here are fabricated from amorphous \mosi{0.61}{0.39}. Since the film lacks long-range crystalline order, irradiation is not expected to introduce disorder in the same way as in polycrystalline superconductors. However, He-ion irradiation may still modify the local structure, composition, interfaces, and thermal environment of the film. These changes can affect $T_\mathrm{c}$, $R_\mathrm{s}$, $I_\mathrm{sw}$, and the overall detector performance. To investigate whether irradiation-induced changes occur and whether they improve detector performance, we compare non-irradiated devices with detectors exposed to different He-ion fluences $F$. As a first measure of sensitivity, we analyze the detection current \idet and saturation current $I_\mathrm{sat}$, shown in \Cref{fig:countrate_plot}.

\subsection{Detector performance}
\begin{figure}[t]
    \centering
    \includegraphics[width=\columnwidth]{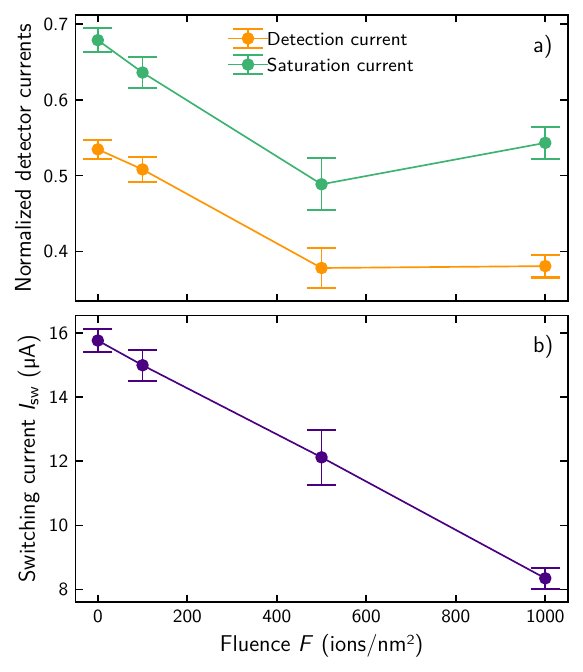}
    \caption{a) \idetnorm and \isatnorm for \SI{780}{\nano \meter} photons at different $F$ normalized to the respective switching current.
    b) Switching current statistics of detector fabricated from \mosi{0.61}{0.39} with \SI{100}{\nano \meter} nanowire width and \SI{4.9}{\nano \meter} thickness, measured at a base temperature of \SI{1}{\kelvin}. The error bars show the standard deviation of \isw across 250 acquisitions per device.}
    \label{fig:detector_currents}
\end{figure}

We measure the bias-current dependent count rate curve and fit the data using a sigmoid function \cite{Autebert2020,Kozorezov2017}. 
In the saturation regime, the count rate does not increase further by increasing the bias current, meaning that all absorbed photons generate a detection event. The saturating quantum efficiency is commonly used as a fingerprint of a highly efficient detector.
As shown in \Cref{fig:countrate_plot}, we define the current at the inflection point, corresponding to \SI{50}{\percent} of the maximum count rate, as the detection current $I_\mathrm{det}$. 
The saturation current is defined as the current at which \SI{95}{\percent} of the saturation count rate is reached.  A detailed explanation of the extraction of \idet can be found in Grotowski \textit{et al.} \cite{Grotowski2026}. 
\figreflet{fig:detector_currents}{a} shows \idetnorm and \isatnorm, as a function of $F$. 
The values are normalized to the mean switching current (averaged over 250 acquisitions) of the respective detector, which are shown in \figreflet{fig:detector_currents}{b}. 
The data shows that for the non-irradiated device \idetnorm is around 0.52 and the count rate saturates at around 0.66. 
Both quantities decrease with increasing irradiation, with the normalized \idetnorm reaching a minimum of 0.36 compared to 0.52 for the non-irradiated detector at \SI{500}{ions \per \nano \meter ^2}. 
At the highest fluence of $F\,= \SI{1000}{ions \per \nano \meter ^2}$ \idetnorm slightly increases to 0.37. \isatnorm follows the same trend, shifted to slightly higher values, where an optimum is more clearly pronounced. For the detector at \SI{1000}{ions \per \nm \squared} we observe a dark count rate below \SI{2}{\hertz} in the saturation plateau and a count rate above \SI{250}{\kilo \hertz} upon illumination with \SI{780}{\nm} photons.
Simultaneously, the switching current decreases almost linearly with increasing $F$. 
\figreflet{fig:detector_currents}{b} shows the switching-current statistics with 250 acquisitions per detector. Overall, the switching currents show only a small spread, with a relative standard deviation of less than \SI{7}{\percent}.

\begin{figure}[h]
    \centering
    \includegraphics[width=\columnwidth]{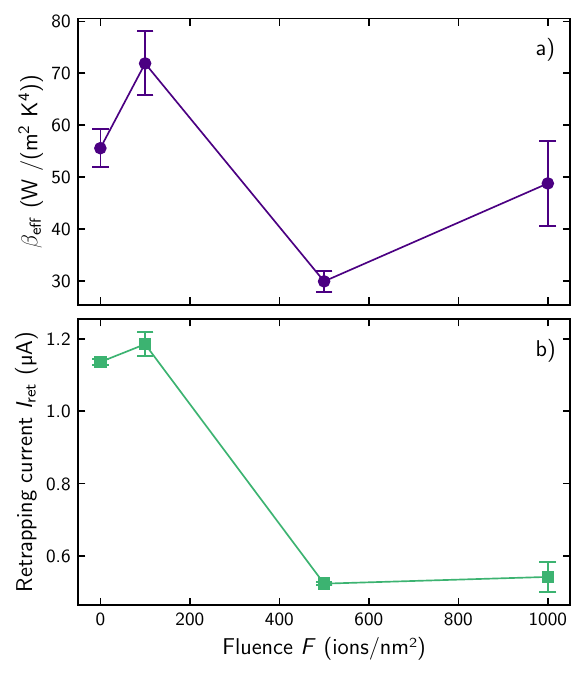}
    \caption{\iret and heat conductance to substrate $\beta_\mathrm{eff}$ at \SI{1}{\kelvin} versus $F$. The decrease with fluence is most apparent for the return current and less conclusive for $\beta_\mathrm{eff}$ featuring a minimum at \SI{500}{ions \per \nano \meter ^2}.}
    \label{fig:retrapping}
\end{figure}

We continue by investigating the origin of the observed sensitivity enhancement. A reduced \csub $\beta_\mathrm{eff}$ can be associated with enhanced detector sensitivity, as a lower thermal coupling to the substrate confines more of the deposited energy within the superconducting film during hotspot formation. \cite{Strohauer2025, Hong2025} Also, improvements in detection efficiency, associated with reduced \csub have been confirmed \cite{Xu2023}. In order to analyze the changes in the \csub, we measure the retrapping current of the detectors, defined as the minimal current that can sustain a resistive region in the nanowire \cite{daneSelfheatingHotspotsSuperconducting2022,
stockhausenAdjustmentSelfheatingLong2012}. As shown in \Cref{fig:retrapping}, the retrapping current \iret, measured at \SI{1}{\kelvin} shows an overall downward trend with increasing ion fluence, reaching approximately half of its initial value. In contrast, the extracted \csub shows no clear systematic dependence on fluence. A slight decrease may be present, especially with the dip for the \SI{500}{ions \per \nano \meter ^2} device, but the variation remains comparable to the experimental uncertainty.

We therefore conclude that the enhanced detector sensitivity is likely dominated by the reduction of the superconducting energy gap due the decrease in \tc (\Cref{sec:transport}), while changes in \csub appear to play only a minor role in the investigated fluence range. 

Overall, He-ion irradiation can improve the photon sensitivity of amorphous \ce{MoSi} SNSPDs, even though the switching current is reduced. The results indicate a clear modification of the detector performance, while the role of changes in thermal conductance remains less conclusive. Therefore, we will focus the further analysis on the irradiation-induced changes on the superconducting thin film. 

\subsection{Transport}
\label{sec:transport}
In the following, the transport properties of the thin films are evaluated as well as the underlying ion induced changes in the thin film.
We start with \tc for two different thin films with Mo-concentrations of 0.61 and 0.74 that will be referred to as Film~A and Film~B, respectively. 

\begin{figure}[h]
    \centering
    \includegraphics[width=\columnwidth]{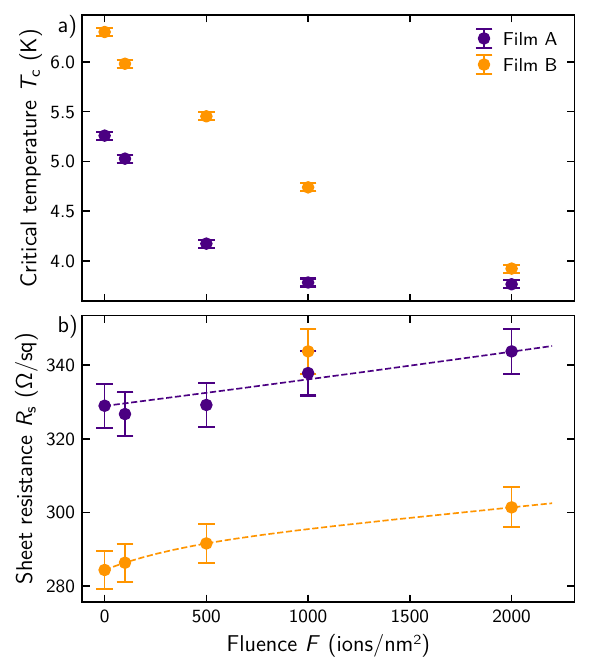}
    \caption{Electrical transport measurements determined using CL structures. \tc at different $F$ for Film~A and Film~B. Both curves show the highest \tc for the non-irradiated CL and a decreasing trend for higher $F$. b) \rs for different $F$ for Film~A and Film~B. Both curves are fitted with an empirical model \cite{Strohauer2023} and show an increase compared to the non-irradiated reference. The resistance increased only 4.5\% for film A and 6\% for film B comparing highest to lowest fluence. The fitting parameters can be found in \Cref{tab:resistance_fit_parameters}.}
    \label{fig:transport_vs_fluence}
\end{figure}

The \tc of both films decreases strongly with increasing $F$, as shown in \figreflet{fig:transport_vs_fluence}{a}. 
For Film~A, \tc decreases from \SI{5.3}{\kelvin}, with the reduction becoming weaker at higher fluences and approaches a saturation value around \SI{3.8}{\kelvin}.
Film~B shows a similar decreasing trend from \SI{6.3}{\kelvin} to \SI{3.9}{\kelvin} but no saturating behavior is observed.

The second metric under investigation is the room-temperature sheet resistance \rs as shown in \figreflet{fig:transport_vs_fluence}{b}. 
The unirradiated film shows a sheet resistance of \SI{328}{\ohm \per sq} and \SI{284}{\ohm \per sq} for Film~A and Film~B, respectively. 
It increases only moderately with irradiation for both thin-film configurations by approximately \SI{5}{\percent} from the initial non-irradiated sample to the highest dose of \SI{2000}{ions \per \nano \meter^2}. 
The data point of Film~B at \SI{1000}{ions \per \nano \meter ^2} is most likely an outlier due to a defect in one of the CL arms and therefore shows an overestimated value of \SI{345}{\ohm \per sq}, which is why we exclude this data point in the further analysis. 
A related study by Kohop{\"a}{\"a} \textit{et al.} \cite{Kohopaa2024} investigated ion irradiation of \ce{MoSi} from separate \ce{Mo} and \ce{Si} layers irradiated with heavier \ce{Ga} and \ce{Ar} ions. In their study, they attributed the increase in \tc partly to irradiation-enhanced intermixing. The different initial film configuration limits direct comparison with our initially homogeneous amorphous \ce{MoSi} films, however, our observation is in agreement with the reported increases in \rs. The quasiparticle diffusivity seems rather unchanged and can be assumed constant for both films and with regards to fluence $F$ with an approximate value of $D=\SI{0.45}{\cm \squared / \second}$ (see \Cref{appendix:diff}).

Next, we want to model the evolution of \rs due to an increase in defect density and an effective thickness reduction with the He-ion fluence $F$. We analyze our results by fitting the data with the empirical defect model established by Strohauer \textit{et al.} \cite{Strohauer2023}. 
The increasing \rs is given by
\begin{equation}
    R_\mathrm{s}(F)= \frac{1}{v_\mathrm{D}}
    \left(1-
    (1-n_{\mathrm{D},0} \, v_\mathrm{D}) \, 
    e^{-\eta v_\mathrm{D}^{2/3} F}
    \right)
    \frac{a_{d_0}}{d_0 - r_\mathrm{r} F},
    \label{eq:R_sheet_phyiscal_fit}
\end{equation}
considering the amount of defect containing clusters of volume $v_\mathrm{D}$ created with an efficiency $\eta$. Here, $n_{\mathrm{D},0}$ is the initial defect cluster density of the non-irradiated film, which can be understood as a disorder estimation, ranging from 0 (crystalline) to 1 (amorphous). 
Additionally, the model accounts for a thickness reduction due to surface sputtering or other processes by the last term. 
$d_0$ denotes the initial film thickness, $r_\mathrm{r}$ the thickness reduction rate and $a_{\mathrm{d}_0}$ a film-thickness dependent constant, related to the resistivity of the pristine film.

Intuitively, the model assumes that irradiation progressively converts initially unaffected regions of the film into defect-containing regions. As the number of unaffected regions decreases, further defect generation becomes less likely and the defect contribution to the resistance gradually saturates. In contrast, the effective film thickness can continue to decrease with fluence, causing the sheet resistance to increase even after the defect density approaches saturation.

We find that despite the small absolute change, the overall increase remains consistent with the model. 
The fit yields an initial internal defect fraction $n_{\mathrm{D,0}}v_\mathrm{D} = 0.99 \pm 0.02$ for Film~A and $n_{\mathrm{D,0}}v_\mathrm{D} = 0.97 \pm 0.02$ for Film~B.
Such high initial values of the defect density are consistent with the amorphous nature of \ce{MoSi}, which already contains a high degree of structural disorder prior to irradiation. 
Within the empirical model, this leaves only a small contribution from the exponentially saturating defect-generation term, making the defect generation efficiency parameter less meaningful.

\begin{table}
    \centering
    \begin{tabular}{lcccc}
        \hline
        Film &
        $\frac{a_{d_0}}{v_D}$&
        $n_{D,0}v_D$ &
        $\eta v_D^{2/3}$&
        $r_r$\\
        &
        [\SI{}{\ohm\,\nm}] &
         &
         [\SI{}{\nm^3\,ion^{-1}}]&
         [\SI{}{\nm \squared\,ion^{-1}}]\\
        \hline
        A & 1644.49 & 0.9936 & 1 & $10.7\times10^{-5}$ \\
        B & 1453.13 & 0.9785 & $2.66\times10^{-3}$ & $8.93\times10^{-5}$ \\
        \hline
    \end{tabular}
    \caption{Fit parameters of the empirical resistance model. $\eta v_D^{2/3}$ is of limited meaning since the film is amorphous, therefore the exponentially modified contribution is very small. This explains the big difference in defect generation efficiency, especially for film~A which has basically no exponential contribution. $\frac{a_{d_0}}{v_D}$, related to the initial resistance and also the thickness reduction rate $r_r$ are similar for both films. Relative uncertainities are below \SI{2}{\percent} with the exception of $r_r$ showing an uncertainty of \SI{48}{\percent}.}
    \label{tab:resistance_fit_parameters}
\end{table}
This leads to the conclusion that the irradiation-induced resistance increase is dominated by the contribution associated with the effective film thickness. 
Overall, He-ion irradiation affects \tc of \ce{MoSi} considerably stronger than $R_\mathrm{s}$. 
The small increase in \rs may be due to the amorphous nature of \ce{MoSi} while for the \tc reduction other contributors play a bigger role. Using the Simonin model \cite{Simonin1986, Banerjee2017} (see \Cref{appendix:simonin}), we can exclude thickness decrease as the main reason of \tc reduction. 

Lastly, it shall be noted that \tc varies with stoichiometry \cite{Grotowski2025}. Since the material is capped with \SI{4}{\nano \meter} of \ce{Si} (with a native oxide formed on top), the interaction of ion irradiation with the capping layer and its potential influence on the superconductor also need to be considered. In particular, recoiled \ce{Si} or \ce{O} atoms from the capping layer can enter the MoSi and change its local stoichiometry, which may in turn affect $T_\mathrm{c}$. We quantify the influence of these two contributions using simulations in the next section.

\subsection{SRIM recoil simulation}
\subsubsection{Simulation setup}
\begin{figure}[t]
    \centering
    \includegraphics[width=\columnwidth]{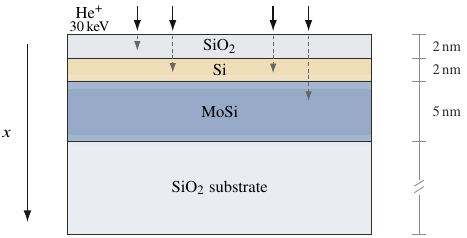}
    \caption{Stack used for the SRIM simulations. \SI{2}{\nano\meter} of the \ce{Si} capping layer is oxidized, while a \SI{5}{\nano\meter} \ce{MoSi} layer remains on top of the \ce{SiO2} substrate.}
    \label{fig:SRIM_stack}
\end{figure}

In this last section, we investigate the origin of our observations with simulations of the Stopping and Range of Ions in Matter (SRIM), which is based on the Monte-Carlo method \cite{Ziegler2010}. 
In particular, He-ion irradiation can induce intermixing at both interfaces of the superconducting layer. \ce{Si} and \ce{O} from the capping and substrate can be recoiled into the \ce{MoSi}, while \ce{Mo} and \ce{Si} from the superconducting layer can be displaced into the substrate. This redistribution can modify the local stoichiometry and effective composition of the superconducting film. We discuss the recoil profile obtained from simulations and how these recoils affect the superconductor and its properties.
The material stack for the SRIM simulation is shown in \Cref{fig:SRIM_stack} and consists of \SI{2}{\nano \meter} oxidized \ce{SiO2} \cite{Klein2026, Bohling2016, Fritzsche1979} followed by the remaining \SI{2}{\nano \meter} of the \ce{Si} capping layer, \SI{5}{\nano \meter} of MoSi and lastly, the \ce{SiO2} substrate. 

\subsubsection{Recoil profile}
\begin{figure}[t]
    \centering
    \includegraphics[width=\columnwidth]{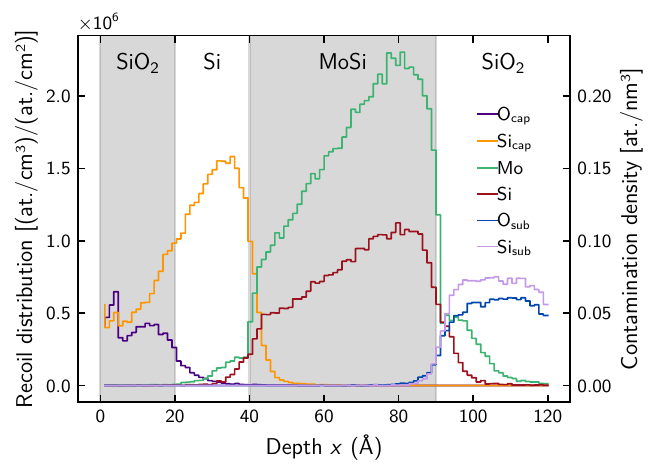}
    \caption{SRIM recoil distribution of Si$_\text{cap}$ (orange) native to the capping layer, \ce{Mo} (green) and \ce{Si} (red) are within the superconducting thin film and O$_\text{sub}$ (blue) and Si$_\text{sub}$ (lavender) are native to the substrate layer. The twin y-axis shows the contamination density after multiplication by a reference ion fluence of $F = \SI{1}{ion \per \nano \meter^2}$.}
    \label{fig:SRIM_rec_profile}
\end{figure}

\Cref{fig:SRIM_rec_profile} shows the recoil distribution profile as a function of depth, ranging from the top at \SI{0}{\angstrom} to \SI{120}{\angstrom}. The recoil profile describes where atoms of each species are located after being displaced by the incident ions. The unit \si{(atoms/cm^3) / (atoms/cm^2)} is practical since it allows to calculate the corresponding contamination density by multiplication with a reference ion fluence of $F=\SI{1}{ion \per \nano \meter ^2}$, shown on the second y-axis. 
We distinguish the origin of the recoiled atoms: O$_\text{cap}$ (blue) is native to the first layer, Si$_\text{cap}$ (orange) to the first two layers, \ce{Mo} (green) and \ce{Si} (red) to the third layer while O$_\text{sub}$ (purple) and Si$_\text{sub}$ (brown) originate from the substrate layer. This distinction allows to track atoms of the same species from different layers separately.
The recoil distance depends on the transferred energy, atomic masses, and material properties such as density and binding energy \cite{Ziegler2010}.
Most recoiled atoms remain within the layer in which they originate, most prominently for \ce{Si} in the capping layer and \ce{Mo} and \ce{Si} in the superconductor. However, some atoms receive sufficient energy to cross a layer boundary, leading to intermixing and thus contamination of the adjacent layer. In some cases, atoms travel even further, such as \ce{O} crossing the pristine \ce{Si} capping layer and reaching the \ce{MoSi} layer. We observe that the amount of recoils decays approximately exponentially with distance from the interface.
Additionally, we see that atoms can also recoil towards the surface, opposite to the direction of the incident ion beam. However, the occurrence is much smaller and recoils towards deeper layers are more probable. In the superconductor, most recoils stay within the layer with only a few atoms being removed into adjacent layers. The most recoils are \ce{Mo} atoms due to the abundance of \ce{Mo} in the \mosi{0.61}{0.39} film.

\subsubsection{Contamination}
So far, we have established the recoil profiles, where atoms are distinguished by both element and layer of origin. The contamination fraction $\varphi_\text{cont}$, defined as the ratio of foreign atoms to host atoms, can then be obtained by dividing the contamination density by the atomic density of the respective layer.
In our further analysis, we want to quantify the contamination in the superconducting layer.
Following the previously observed exponential decrease of recoils with distance from an interface, we can fit the individual profiles with an exponential function. The detailed fit procedure can be found in \Cref{appendix:fit}. 
The contributions \ce{O} and \ce{Si} originating from different layers are combined into a total \ce{O} and \ce{Si} contamination profile respectively. \ce{Si} from the \ce{MoSi} layer is not considered as contamination. Multiplication by the He-ion fluence relative to the reference fluence of \SI{1}{ion \per \nano \meter^2} then yields the fluence-dependent contamination profile $c(x,F)$.

\subsubsection{Oxidation model}
\label{sec:oxidation_model}
We start by analyzing the influence of \ce{O} recoils and consider \ce{Si} separately in the next section. 
To describe the irradiation-induced changes of the superconducting film, we introduce a fluence-dependent oxidation model that maps oxygen intermixing at the interfaces onto a reduced effective superconducting thickness. 
The model assumes that oxygen incorporation and subsequent oxidation of \ce{MoSi} produces an insulating oxide. This is proposed as the dominant contribution to the reduction of the electrically active thickness.\\
The model treats the superconducting film as discrete depth segments $\Delta d$, which can be partially or fully oxidized depending on the local oxygen contamination. 
For each depth segment $\Delta d$, the oxygen contamination $c_\mathrm{O}(x,F)$ is defined as the ratio of incorporated oxygen atoms to the initial number of Mo and Si atoms in that segment. We introduce a phenomenological threshold $c_\mathrm{thr}$ at which the segment is assumed to become fully oxidized and electrically inactive. For example a threshold $c_\mathrm{thr}=1$ means we need one \ce{O} per host atom for full oxidation, $c_\mathrm{thr}=2$ means two \ce{O} per host atom (e.g. to form \ce{SiO2}). Below this threshold, the loss of electrically active material is assumed to scale linearly with the oxygen contamination. This linear relation is a phenomenological approximation and does not describe the microscopic oxidation process.

The electrically active fraction of each segment is therefore defined as
\begin{equation}
f_\mathrm{active}(x,F) = 1-f_\mathrm{O}(x,F)  = 1-
\min\left(1,\frac{c_\mathrm{O}(x,F)}{c_\mathrm{thr}}\right),
\end{equation}
where $c_\mathrm{O}(x,F)$ is the local oxygen contamination at depth $x$ and ion fluence $F$. Consequently, a segment at or above $c_\mathrm{thr}$ contributes no electrically active thickness. A segment at \SI{50}{\percent} of $c_\mathrm{thr}$ contributes \SI{50}{\percent} to the electrically active thickness. 
The entire remaining thickness of the superconducting layer $d_\mathrm{eff}$ is obtained by integrating the electrically active fraction over the depth $x$ of the superconducting film, 
\begin{equation}
    d_\mathrm{eff}(F) = \int_{0}^{d} \left[ 1- 
    \min\left(1,\frac{c_\mathrm{O}(x,F)}{c_\mathrm{thr}}\right) \right]\,dx .
    \label{thickness_integral_contamination}
\end{equation}

This results in a thickness reduction of the thin film due to oxidation in an almost linear way with a total decrease of less than \SI{5}{\percent} at \SI{2000}{ions \per \nm^2}. Interestingly, the surface sputtering extracted from SRIM is negligible in this scenario since its rate is below 0.06 atoms per He-ion. This results in a thickness reduction of \SI{0.65}{\nm} at a high dose of \SI{2000}{ions \per \nano \meter^2} and therefore only affecting the very top of the capping layer and not affecting the superconducting layer.

\begin{figure}[t]
    \centering
    \includegraphics[width=\columnwidth]{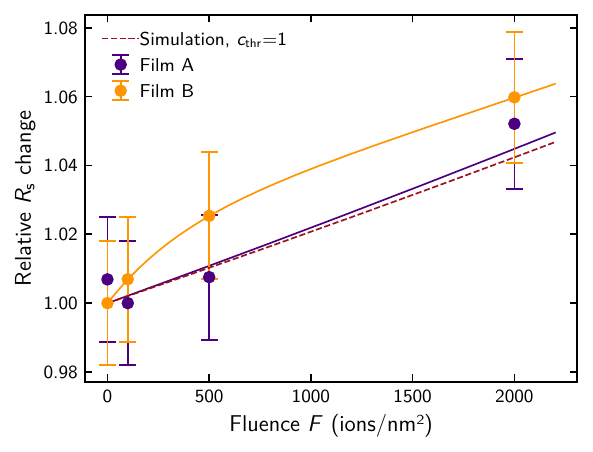}
    \caption{Dynamic oxidation model demonstrated for the relative sheet resistance increase with ion fluence $F$. \rs and fits using the empirical model obtained from \Cref{fig:transport_vs_fluence} show good agreement with the trend from SRIM simulation and the dynamic oxidation model, especially for Film~A.}
    \label{fig:dynOx}
\end{figure}

The sheet resistance \rs is not only dependent on the resistivity, which itself is influenced by defects and crystal structure, but also on the thickness of the material. Therefore, a reduction in thickness will result in an increase in $R_\mathrm{s}$, according to

\begin{equation}
    R_s(F) = R_{s0} \cdot \frac{d_0}{d_\mathrm{eff}(F)}
    \label{eq:ox_model}
\end{equation}
with $R_{s0}$ and $d_0$ being the initial sheet resistance and thickness and $d_\mathrm{eff}(f)$ the fluence dependent electrically active thickness. 

We compare the relative \rs increase obtained from simulation and measurement in \Cref{fig:dynOx} by normalizing \rs to its minimum value and using $c_\mathrm{thr}=1$. Overall, \rs increases by approximately \SI{5}{\percent} between \SI{0}{ions \per \nano \meter^2} and \SI{2000}{ions \per \nano \meter^2}. Film~B shows a larger resistance increase due to the stronger contribution of the exponential term in the empirical defect model. In contrast, Film~A has only a small exponential contribution and therefore agrees well with the prediction of the dynamic oxidation model. A slightly lower value of $c_\mathrm{thr}=0.95$ would further improve the agreement with Film~A.
Although the measured and simulated trends appear nearly linear, the underlying model is not strictly linear. Oxygen-induced degradation reduces the effective superconducting thickness $d_\mathrm{eff}$, and the corresponding sheet resistance is estimated from \Cref{eq:ox_model}. This results in a reciprocal dependence of the form $a/(a-x)$, which appears approximately linear as long as the effective thickness reduction remains small. We want to emphasize that $c_\mathrm{thr}$ is a phenomenological parameter rather than a fixed physical constant.

\subsubsection{Stoichiometry changes}
In the recoil distribution we observed, that atoms can recoil in different layers from where they originate like the \ce{Si} contamination of the \ce{MoSi} layer. 
A direct consequence of this is an associated stoichiometry change, which we investigate in the following. 
\begin{figure}[h]
    \centering
    \includegraphics[width=\columnwidth]{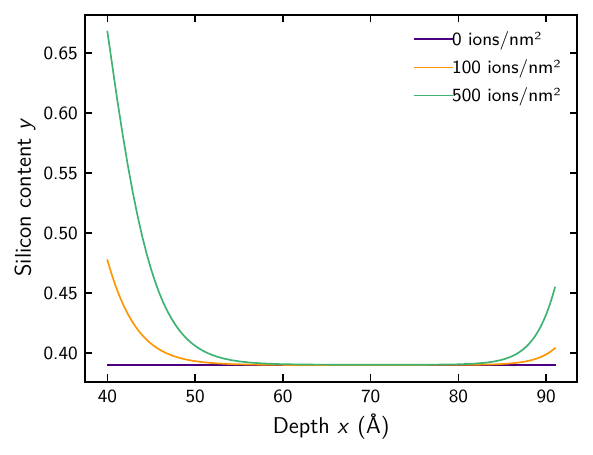}
    \caption{Silicon stoichiometry versus film depth $x$. Silicon content $y$ in \ce{Mo_{1-y}Si_y} increases from the initial \mosi{0.61}{0.39}, especially at the interface and the unperturbed central part shrinks with increasing $F$.}
    \label{fig:SRIM_SI_stoich}
\end{figure}
\Cref{fig:SRIM_SI_stoich} shows the stoichiometry in the superconductor layer as a function of depth for different ion fluence $F$. 
The largest \ce{Si} incorporation occurs close to the layer boundaries and decreases rapidly towards the center of the \ce{MoSi} layer. In particular, the upper interface shows a considerably stronger contamination than the interface towards the substrate. Therefore, most of the additional \ce{Si} is concentrated within approximately the first \SI{10}{\angstrom} of the superconducting layer. In comparison, the center of the film remains only weakly contaminated. With increasing $F$, the contamination increases and extends further into the layer, reducing the thickness of the weakly affected central region. For \SI{500}{ions \per \nano \meter^2} the \ce{Si} content at the interface is almost doubled according to the simulation but this may differ from observations in the experiment. SRIM simulates implantation into an unchanged target and does not account for dynamic changes in density or composition in the layer. Therefore, the exact stoichiometry close to the interfaces become increasingly uncertain at high ion fluence.

The resulting change in local stoichiometry is expected to influence the superconducting and normal-state transport properties of the film. Previous studies of \ce{MoSi} with varying composition \cite{Grotowski2026, Edelstein, Zhang2021, Erbe2024} show that an increased \ce{Si} content can be associated with an increased \rs and a reduced $T_\mathrm{c}$. The \ce{Si} incorporation predicted here provides an additional mechanism that can contribute to the irradiation-induced degradation of superconductivity. However, the spatially inhomogeneous composition and the simultaneous redistribution of several atomic species make it difficult to quantitatively predict the resulting transport properties.

In addition to the incorporation of atoms from adjacent layers, native \ce{Mo} and \ce{Si} atoms are themselves recoiled out of the superconducting layer. The removal of \ce{Mo} can further shift the local composition towards a Si-rich material, whereas the simultaneous removal of native \ce{Si} partially counteracts the incorporation of additional \ce{Si}. Furthermore, \ce{Mo} recoiling into the adjacent layers causes the interfaces to become intermixed, making the boundary of the superconducting layer increasingly diffuse. Since these exchange processes modify both the local stoichiometry and the effective layer boundaries, their individual contributions to the reduction of \tc cannot be separated quantitatively within this model and are beyond the scope of this work. However, the analysis shows that the observed \tc suppression cannot be explained by thickness reduction alone (see \Cref{sec:oxidation_model}).

\section{Conclusion and outlook}
In summary, we demonstrated that helium ion irradiation can enhance the intrinsic photon sensitivity of amorphous \ce{MoSi} SNSPDs. The improvement is observed in the bias-current-dependent count-rate characteristics, where both the normalized detection current \idetnorm and saturation current \isatnorm decrease with increasing fluence. The relative saturation plateau increases from 35\% to 45\%.
Although \ce{MoSi} is amorphous and therefore lacks a crystalline structure that can be progressively disordered by irradiation, its superconducting and transport properties are strongly modified. The critical temperature \tc is reduced by approximately \SIrange{30}{40}{\percent}, whereas the room-temperature sheet resistance increases by only about \SI{5}{\percent}. The switching current decreases accordingly, while the \csub shows only a small variation rather than a systematic reduction. Therefore, the enhanced detector sensitivity appears to be dominated by the suppression of superconductivity and the associated reduction of the superconducting energy gap, rather than by changes in the thermal coupling to the substrate.

SRIM simulations provide further insight into the microscopic origin of the irradiation-induced changes and identify oxidation and intermixing as possible contributors. Oxygen incorporation is mapped onto a gradual reduction of the electrically active film thickness using a phenomenological oxidation model, while intermixing of \ce{Si} and \ce{O} modifies the local \ce{MoSi} stoichiometry. The resulting resistance increase shows good agreement of measurement and simulation, supporting the phenomenological oxidation model.

Overall, these results show that helium ion irradiation is not limited to modifying crystalline superconductors but can also be used to tune amorphous silicide-based SNSPDs. The strong enhancement of photon sensitivity makes irradiation particularly promising for extending the detection range towards longer wavelengths, also for materials that already exhibit excellent long wavelength sensitivity. Further analysis of intermixing, defect generation, and interface changes is crucial for a more targeted optimization of ion irradiation techniques. In particular, a position-resolved analysis of irradiation-induced stoichiometric changes could provide valuable insight into the underlying modification mechanisms.

\section*{Funding acknowledgments}
We gratefully acknowledge financial support from the German Federal Ministry of Research Technology and Space (BMFTR) via the funding program “Quantum technologies – from basic research to market” (projects PhotonQ (13N15760), QPIS.2 (16KISQ172), FgED (13N17529), SAEQS (3N16760), SPINNING (13N16214) and QRN (16KIS2197)), as well as from the German Research Foundation (DFG) under Germany’s Excellence Strategy EXC-2111 (390814868) and projects INST 95/1720-1 (MQCL) and PQET (INST 95/1654-1). This research is also supported by the ‘Munich Quantum Valley”, which is supported by the Bavarian state government with funds from the “Hightech Agenda Bayern Plus”.

\section*{Author Declarations}
\subsection*{Conflict of interest}
The authors have no conflicts to disclose.

\subsection*{Data availability}
The data supporting the findings of this study are available from the corresponding author upon reasonable request.

\section{Appendix}
\subsection{Quasiparticle diffusivity}
\label{appendix:diff}
\begin{figure}[h]
    \centering
    \includegraphics[width=\columnwidth]{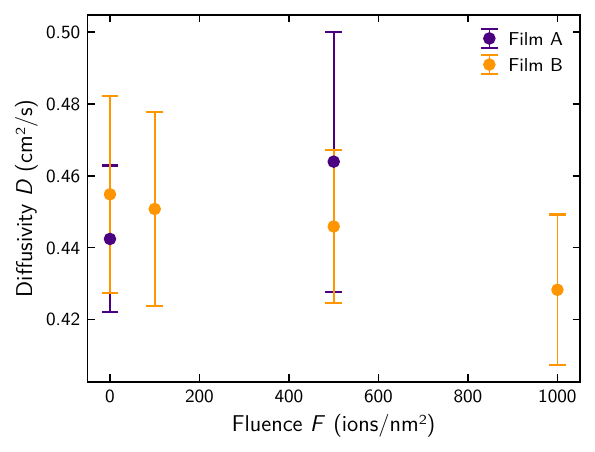}
    \caption{Quasiparticle diffusivity of Film~A and B only decreases very slightly with $F$, while the decrease is within the measurement uncertainty. $D$ can be assumed to be constant for both films and with respect to $F$.}
    \label{fig:diffusivity}
\end{figure}
The quasiparticle diffusivity \cite{Catelani2022} as shown in \Cref{fig:diffusivity} barely changes with irradiation. The uncertainty is larger than the change with fluence, however a slight decrease with $F$ can be observed. For practical reasons, $D$ can be assumed constant with $F$ at a value of $D_\mathrm{A}=$ \SI{0.45}{\cm^2 \per \s} and $D_\mathrm{B}=$ \SI{0.44}{\cm^2 \per \s}.

\subsection{Simonin model}
\label{appendix:simonin}
\begin{figure}
    \centering
    \includegraphics[width=\columnwidth]{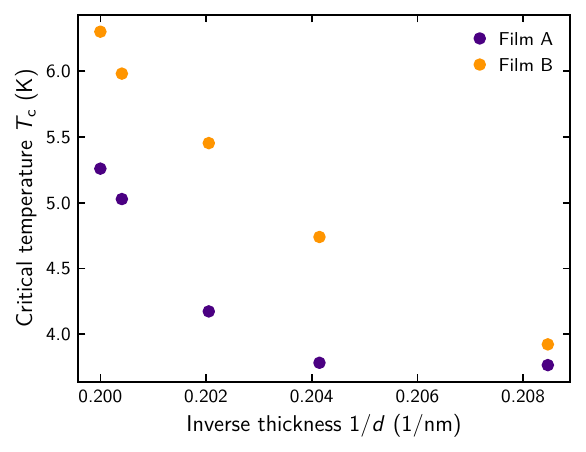}
    \caption{\tc versus inverse film thickness, obtained from SRIM simulation using our proposed dynamic oxidation model. The lack of linearity suggests that the reduction in \tc is not only due to the effective thickness reduction caused by \ce{O} intermixing.}
    \label{fig:simonin}
\end{figure}

According to the phenomenological model proposed by Simonin \cite{Simonin1986, Banerjee2017}, \tc should depend linearly on $1/d$, with the intercept at $1/d=0$ corresponding to the bulk \tc and the limit \tc=0 defining the critical thickness for superconductivity. \Cref{fig:simonin} shows the critical temperature as a function of inverse effective film thickness, using the thickness reduction obtained from the dynamic oxidation model. However, no linear dependence is observed. The measured change in \tc therefore cannot be explained by the predicted thickness reduction alone, indicating that irradiation additionally modifies the intrinsic superconducting properties of the film.

\subsection{Fitting of contamination profiles}
\label{appendix:fit}
The contamination fraction $\varphi_\text{cont}$ for a single species (distinct element from a certain layer) is approximated with an exponential function
\begin{equation}
    \varphi_\text{cont} = A \cdot e^{-B\cdot x} + C
\end{equation}

\begin{figure}
    \centering
    \includegraphics[width=\columnwidth]{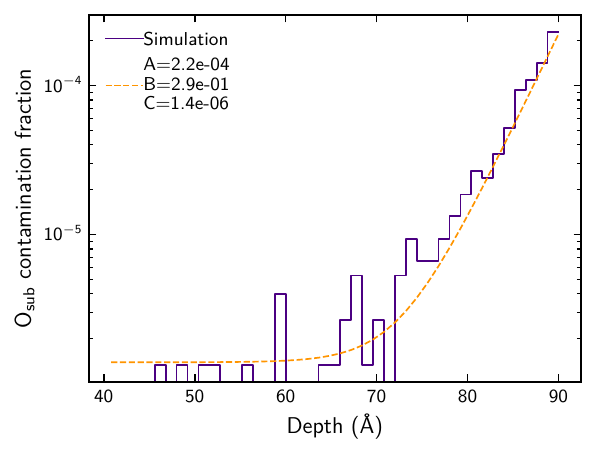}
    \caption{Exemplary fit of oxygen contamination fraction $\varphi_\text{cont}$ obtained from simulation using an exponential function with constant background. The fits are used to calculate the contamination profile for \ce{O} and \ce{Si} and determine their influence on the superconductor.}
    \label{fig:example_cont_fit_oxygen}
\end{figure}

Here, $A$ describes the magnitude of the contamination close to the interface, $B$ the decay with distance, and $C$ a constant background contribution. The coordinate $x$ represents the distance from the respective interface into the \ce{MoSi} layer. An example of the fit is shown in \Cref{fig:example_cont_fit_oxygen}. Using analytical fit functions instead of the discrete SRIM data simplifies the subsequent calculation and allows the contamination profile to be evaluated continuously throughout the superconducting layer.

\bibliographystyle{unsrtnat}
\bibliography{references}
\end{document}